\documentclass[a4paper,fleqn,authoryear]{cas-dc}

\usepackage[numbers]{natbib}
\usepackage{float}
\usepackage{threeparttable}

\newcommand{\rqone}{What ideas do K-12 educators have for learning opportunities using MLLMs?}
\newcommand{\rqtwo}{What issues emerged through educators’ experiences prototyping with an MLLM, and how do these challenges shape educators’ views around anticipated student use?}

\def\tsc#1{\csdef{#1}{\textsc{\lowercase{#1}}\xspace}}
\tsc{WGM}
\tsc{QE}
\tsc{EP}
\tsc{PMS}
\tsc{BEC}
\tsc{DE}

\begin{document}
\let\WriteBookmarks\relax
\def\floatpagepagefraction{1}
\def\textpagefraction{.001}
\shorttitle{Beyond Text: Probing K-12 Educators' Perspectives on Learning Opportunities with Multimodal Large Language Models}
\shortauthors{Tseng et al.}

\title [mode = title]{Beyond Text: Probing K-12 Educators' Perspectives on Learning Opportunities with Multimodal Large Language Models}                      








\author[1]{Tiffany Tseng}
\ead{ttseng@barnard.edu}
\cormark[1]

\author[1]{Katelyn Lam}

\author[2]{Tiffany Lin Fu}

\author[1]{Alekhya Maram}

\affiliation[1]{organization={Barnard College},
city={New York},
state={NY},
country={USA}}

\affiliation[2]{organization={Columbia University},
city={New York},
state={NY},
country={USA}}

\cortext[cor1]{Corresponding author}


\begin{abstract}
Multimodal Large Language Models (MLLMs) are beginning to enable new user experiences from generated content across a range of media, including images, text, speech, and video. These capabilities have the potential to enrich learning by enabling users to interact with information using a variety of modalities, but little is known about how \textit{educators} envision how MLLMs might shape the future of learning, what challenges they encounter when using these models, and what practical needs should be considered for future implementation in educational contexts. We investigated educator perspectives through workshops with 12 K-12 educators, where participants brainstormed learning opportunities, discussed practical concerns, and prototyped MLLM learning applications using Claude 3.5 and its Artifacts feature. Through this work, we uncover how educators imagined MLLMs as a way for themselves and their students to author multimedia content, and how this could provide a pathway to support learning through iterative design. At the same time, educators anticipated challenges with younger students' ability to evaluate and refine model output to better meet their design goals. We end with implications for designing with and for MLLMs in future learning experiences.
\end{abstract}




\maketitle

\section{Introduction}
\label{sec:intro}

The widescale release and public adoption of Large Language Model (LLM) tools like ChatGPT has the potential to enable powerful opportunities for personalized learning. Because these models have been trained on vast amounts of data, they can provide on-demand explanations across a wide range of domains, which has naturally led researchers, policymakers, and education stakeholders to consider how this technology will impact the future of education. 

Many AI companies have begun to preview \textit{multimodal LLMs} (MLLMs), or models that can interpret and generate content across multiple modalities such as image, audio, video, and text, providing opportunities to move beyond text-based explanations. Videos demonstrating the capabilities of MLLMs showcase educational use cases such as supporting language learning (pointing a smartphone at objects and having the model verbally supply the name of items in a foreign language \cite{gpt4o}) or providing personalized feedback on hand-drawn sketches using live camera input \cite{khanmigo}. These corporate-driven visions of AI-supported learning accompany researcher-led position papers about how LLMs might transform education \cite{kasneci2023chatgpt, kuchemann2024large, gan2023large}; yet, a missing voice is that of \textit{educators} and how they imagine ways MLLMs can be effectively and responsibly used for learning. Incorporating educator perspectives may illuminate unrealized opportunities for applying MLLMs towards enriching learning experiences, particularly for diverse learners and environments \cite{cukurova2018measuring, luckin2019designing, holmes2019artificial, qin2020understanding}. Studying educators' approaches to actively using MLLMs may also surface usability or misalignment issues with current models, which could inform the future development of these technologies to better support educational applications \cite{10.1145/3564721.3564738, kizilcec2024advance}.

To understand the ideas, needs, and challenges educators anticipate around learning experiences using MLLMs, we conducted workshops with 12 K-12 educators in which they reacted to demonstration videos of unreleased MLLMs, brainstormed educational opportunities using MLLMs, and engaged in hands-on prototyping of learning activities using Claude 3.5, a multimodal model which supports text and image input and code-based interactive applications through a feature called Artifacts. Through these workshops, we examined the following research questions:

\begin{itemize}
    \item \textbf{RQ1}: \rqone
    \item \textbf{RQ2}: \rqtwo
\end{itemize}

We found that teachers primarily imagined MLLMs being used to create multimedia simulations and demos, implement personalization and accommodations, and support student research, with a focus on multimodal output rather than input. Prototyping with Claude 3.5, teachers were able to generate code-based interactive applications enabled teachers to create experiences similar to those that they envisioned when designing with a multimodal model that could support multimedia visual output. This finding suggests that aspects of multimodal output may already be achievable through code generation, even before multimodal modalities are made widely available in commercial end-user tools.

Further, we found a distinction between teachers who envisioned designing with MLLMs to build learning tools for student use, versus those who imagined students using MLLMs directly to support exploratory research. This division stemmed from concerns that younger students with varying literacy and communication skills would need substantial support to evaluate model output and persist through the iterative prompting often required to generate content aligned with their goals. Yet, for students who may interact with the model directly, teachers anticipated that the iterative process of refining and debugging generated content may provide richer opportunities for learning distinct from one-shot question-answer exchanges commonly shown in tech demo videos. Overall, we contribute a first look into end-user educator perspectives into MLLMs to inform the design of future MLLM-driven media-rich learning experiences.
\section{Related Work}
\label{sec:related-work}

The application of artificial intelligence to education more broadly has been explored in decades of work, historically stemming from cognitive tutors and adaptive learning systems \cite{anderson1995cognitive, marouf2024enhancing, ritter2007cognitive} and more recently spurred by the introduction of commercial LLMs, particularly ChatGPT in 2022. AI Education efforts range from student- \cite{10.1145/3631802.3631829} to teacher-focused \cite{holmes2022state} and include both applications of AI technologies to support end-users, along with the development of frameworks, tools, and curriculum to foster AI literacy skills \cite{long2020ai, tseng2024co, lee2021developing}. Proponents of utilizing AI technologies for education highlight potential benefits such as enabling customization and personalization for a range of student learning needs \cite{akanzire2025generative} and supporting students' problem solving abilities \cite{guo2025can}; at the same time, there are many concerns over issues such as model bias, risk of student overreliance, and lack of effective policy and infrastructure to support equitable use \cite{bewersdorff2025taking, kuchemann2024large, futterer2023chatgpt, mintz2023artificial}. Some AI companies are beginning to release analyses of how their tools are being used in educational contexts but focus specifically on higher-education, such as Anthropic's Education Report on how college students use Claude \cite{handa2025education} and a companion report on how Claude models are being used by university professors \cite{benthand2025education}. As pointed out in recent work exploring educators use of ChatGPT, K-12 teachers may have unique needs compared to university faculty because they iterate more regularly on lesson plans \cite{keppler2025making}, highlighting the importance of research studying K-12 educator use of AI technologies.

As our work centers on educator ideas for learning applications of MLLMs, we focus on related work on teacher perspectives on AI and early efforts to explore the application of MLLMs for education.

\subsection{Teacher Perspectives on AI}
Recent work soliciting teacher perspectives on AI technologies have found that educators see themselves as likely to integrate AI into their teaching \cite{kaplan2023generative} to support their productivity and student assessment \cite{celik2022promises, ravi2025co, hays2024chatgpt}. Additionally, there is a growing interest in supporting student learning, such as the development of higher-order thinking skills \cite{braaten}. In the context of literacy education, prior work found that teachers see opportunities for LLMs to provide differentiated instruction by customizing content to students' reading levels, providing direct feedback, using culturally relevant examples, and helping students brainstorm ideas for their own stories \cite{han2024teachers}. \citet{delaney2025teaching} describe how researcher and educator co-design can support teachers integrating GenAI into their own lessons. Major concerns raised by teachers include challenges around attribution and distinguishing between LLM-generated and student-created content, overreliance on AI tools and how this might negatively impact student creativity, and how use of chatbots might take away from traditionally social interactions between students and teachers or students and their peers \cite{kasneci2023chatgpt, han2024teachers, ravi2025co, dangol2026relief, xiao2026teachers}. 

Several studies have reported a positive shift in teacher perspectives in response to direct use with AI technologies \cite{klopfer2024generative, nazaretsky2022teachers}. For example \cite{ali2024picture} found teachers had stronger beliefs around the creative potential of generative AI tools after participating in co-design workshops where they tested text-to-image generators. Research on teacher adoption of GenAI use has found that for some educators, using GenAI can offer a playground for expanding the creative possibilities of their practice \cite{dangol2026relief}. 

A distinction we draw between our research and prior work is that we are primarily interested in how teachers might envision the design of custom learning applications with MLLMs for student-use, rather than primarily using LLMs to streamline teacher-oriented work such as administrative tasks, lesson planning, or assessment \cite{ravi2025co, celik2022promises, collie2025teachers, fan2024lessonplanner, goldman2024using, cheah2025integrating, keppler2025making}. Additionally, our approach to understanding educator ideas involves engaging teachers with hands-on prototyping with MLLMs, and to the best of our knowledge, our work is the first to explore K-12 teachers' use of multimodal capabilities with Claude 3.5 and its Artifacts feature to envision future education-oriented MLLM technologies. 

Prior work such as \cite{keppler2025making} examined teachers' prompting strategies when using ChatGPT, identifying four modes: make for me, find for me, jumpstart for me, and iterate with me, all used specifically in the context of preparing content such as quizzes or lesson plans using text-based prompts. In contrast, our work contributes an empirical study of how educator imagine input and output from models beyond text, with the goal of generating student-facing content beyond lesson plans and assessment.

Finally, while a key component to successful adoption of AI educational technologies is teachers' perspectives and engagement with such tools, they have not traditionally been stakeholders in the design of AI-tools \cite{cukurova2018measuring, luckin2019designing, holmes2019artificial, qin2020understanding}, which we address through co-designing and brainstorming with teachers about this emerging technology.

\subsection{Multimodal LLMs and Implications for Education}
Multimodal large language models (MLLMs) are large language models capable of processing and generating content in multiple modalities including text, images, sound, and video. Existing MLLMs frequently focus on a subset of modalities. For example, text-to-image generators like Dall-E and Stable Diffusion were first introduced in 2021 and 2022 respectively and specialize in generating images from text and image input. Text-to-video LLMs such as Runway Gen-4.5 and OpenAI Sora are capable of creating up to 15 seconds of generated video from text prompts or images. For audio, many models utilize speech-to-text transcription for processing of audio data as text, but more recent models like OpenAI's GPT-4o-tts (introduced March 2025) have direct speech-to-speech capabilities. General purpose models like Google's Gemini Ultra combine these modalities, supporting image, video, and audio.

The introduction of MLLMs has captured the public's imagination; for example Google's demo video showcasing Gemini's multimodal features has garnered 3.2 million views on YouTube \cite{gemini}, showing how an agent might fluently respond to verbal requests and perform tasks like creating video-based games\footnote{It was later exposed that the Gemini demo was not truly accurate, leading to skepticism about its true capabilities \cite{techcrunch}}. Similar processing of real-time camera input have been demonstrated by Open AI but has not been publicly released \cite{sixty_minutes}. NotebookLM introduced audio podcasts generation as an alternative modality to text explanations \cite{notebooklm}. These efforts show significant investment in multimodal features to explore their potential for new user experiences.

In response to the rapid introduction of these multimodal capabilities, researchers have begun to envision how they might impact the future of learning. For example, \citet{kuchemann2024large} and \citet{bewersdorff2025taking} offer a range of ideas for learners, teachers, researchers, and developers of educational applications such as enabling customized learning tools to support specific learner needs and disabilities and enhancing student engagement through the use of multimedia content like animation and flow diagrams. These efforts draw from prior work on the value of multimodality for learning more broadly such as the Cognitive Theory of Multimodal Learning \cite{mayer1997multimedia, mayer2002multimedia, mayer2024past}, which argues for the benefits of text alongside images and other multimedia -- additionally, multiple representations afforded by multimodality have been found to effectively foster student motivation and understanding \cite{treagust2008role}. However, these works largely project what researchers believe are opportunities and do not involve end-users like teachers and students, which is essential for successful adoption of these technologies in education settings \cite{cukurova2018measuring, luckin2019designing}. Recent work studying undergraduate students' perspectives \cite{prasad2025exploring} found that students anticipated MLLMs being beneficial for supporting multimedia content creation, enhancing feedback, and providing personalization; however, research is needed to determine alignment with K-12 settings.

Before the introduction of MLLMs, technologies like Google Lens\footnote{https://lens.google/} (first introduced in 2017) utilized computer vision models for identification and explanation of camera input. Prior work examining parents' opinions about using Google Lens with their children found that they perceived the application to be useful for teaching and learning \cite{nguyen2021determinants}; in comparison, MLLMs are less restricted to specific applications (like plant and animal recognition) and can be applied to a wider range of use cases.  

In summary, MLLMs are an emerging technology with only a subset of modalities commercially available to end-users today. Yet, while researchers are already anticipating how MLLMs might impact learning, there is little empirical work that surfaces the perspectives of teachers directly, which may shed light on unexplored learning opportunities and is critical for adoption in educational contexts. Thus, our work provides an early and needed lens into educator views to inform the development and application of MLLMs for education.
\section{Methodology}
\label{sec:methodology}

To explore educational opportunities using MLLMs, we facilitated a series of workshops for K-12 educators to provide their insights and feedback. These workshops were designed to both introduce what MLLMs are and their envisioned capabilities, as well as provide a hands-on experience for teachers to prototype with them to design student-facing learning activities. Our protocol and data collection plan were approved by our college's Institutional Review Board.

\subsection{Benchmarking and Model Selection}
In advance of our workshops, we tested a variety of commercial MLLMs to compare their multimodal features and determine which to use in our workshops. Our benchmarking in October 2024 involved testing OpenAI’s GPT-4o and GPT-4o mini, Anthropic’s Claude 3.5, and Perplexity AI. At the time, all models could support image input, though they had differences in their image generation capabilities; Perplexity and GPT both had more advanced diffusion-based image generation capabilities compared to Claude, which is only capable of generating simple code-based SVGs. None of these tools supported video input or output. All supported some form of dictation (voice-to-text-input), but none could generate audio output, with the exception of OpenAI's mobile application. Finally, all were capable of handling text input and output, but Claude 3.5 was the only one that enabled inline previews of generated code using its Artifacts feature, which allows users to interact with generated code-based applications such as games and websites.

Model outputs and multimodal capabilities were evaluated on a set of prompts (created by our research team) designed to simulate educational use cases involving generating and interpreting images, generating interactive applications, and providing text explanations such as supporting the design of an illustrated storybook from hand-drawn image input and generating interactive math explanations from hand-drawn diagrams. 

Based on these tests, we decided to utilize Claude 3.5 and focus our workshops around text and image input modalities, along with live previews of generated output using Artifacts. Claude Artifacts enables users to publish their artifacts on Claude servers so they are shareable by URL, which we imagined might be especially useful in classroom and group-learning contexts. (At the time of our study in November 2024, ChatGPT did not have a comparable feature for previewing code-based generated output, but it currently supports this capability as of December 2024.)

\subsection{Participant Selection}
We recruited educators through local Slack communities in New York City and email invitations to alumni from our college's professional development program for local teachers. To solicit the expertise of teachers from diverse backgrounds, participants were screened based on the following criteria: (a) type of school or program (both in and out of school), (b) grade levels taught, (c) subjects taught, (d) years of teaching experience, and (e) familiarity with AI technologies.

Table \ref{tab:participants} summarizes the 12 workshop participants we selected from 47 teachers who responded to our interest survey (pseudonyms are used). When choosing these 12 participants, we optimized, based on teachers' stated availability, having teachers from distinct backgrounds (such as grade levels and in-school or out-of-school context) within any given workshop. Similarly, for exclusion criteria, we avoided grouping teachers who represented similar backgrounds (such as teachers from the same school teaching the same subject).

We organized four 3-person workshops over the course of two weekends in November 2024 for a total of 12 participants. These small workshops allowed for close observation, with a member of our research team shadowing each participant to learn about their design process. Small workshops enabled us to facilitate group discussions in which all participants were actively involved and engaged with the research team, providing us with more in depth feedback on their experience.

Our participants included 7 public school teachers, 2 charter school teachers, 1 magnet school teacher, and 2 museum educators teaching across all K-12 grade levels in New York City. Half of the participants taught K-5 elementary grade levels, five taught middle school, and four taught high school. These educators teach subjects like Reading, Writing, Math, and STEM. They identify as Black or African American (42\%), Hispanic or Latinx (33\%), White (25\%), and Asian (8\%). Ten of the teachers were female, and two were male. Overall, they had significant teaching experience, with an average of six or more years of teaching experience, and a range between 1 to 20+ years of experience\footnote{Our recruitment survey asked participants to declare their years of teaching experience from a dropdown menu of ranges, which is why we report average years of experience as a range}. All participants self-reported as novice LLMs users.

\begin{table*}[t]
\begin{center}
    \begin{threeparttable}
  \caption{Educator Participants in Workshops}
  \label{tab:participants}
  \begin{tabular}{lllp{3.25cm}p{5.5cm}}
    \toprule
    Participant & Teaching Context & Grades & Subjects & Primary Workshop Project Created*\\
    \midrule
    Patricia & Public School & K-5 & STEM & Playground physics animations and interactives \\
    Alex & Public School & 9-12 & Earth and Space Science & Interactive map of NYC floods\\
    Mary & Public School & 9-12 & Computer Science & Interactive for exploring triangle congruence\\
    Karen & Public School & 6-8 & English Language Arts (ELA) & Website for exploring refugee crisis \\
    Esther & Charter School & K-5 & Reading, Writing, Math  & Number sequencing game \\
    Michelle & Magnet School & 6-8 & Biology and Physical Science & Human organ game  \\
    Janet & Charter School & 1-5 & Reading, Writing, Math & Interactive reading guide  \\
    Rachel & Museum & K-12 & Horticulture & Interactive garden recipe game  \\
    Jessica & Museum & 9-12 & STEM & Website comparing habitation on different planets \\
    Leah & Public School & K-8 & English, Math, Science, Social Studies  & Turkey drawing game \\
    George & Public School & 9-12 & Math, Multimedia and Robotics & Website for supporting research on the solar system \\
    Sam & Public School & 6-8 & Math & Interactive for exploring linear inequalities \\
  \bottomrule
\end{tabular}
\begin{tablenotes}
    \small
    \item * Some teachers created multiple projects. The \textit{Primary Workshop Project Created} column indicates the project they spent the most time prototyping.
\end{tablenotes}
\end{threeparttable}
\end{center}
\end{table*}

\subsection{Workshop Design}

\begin{figure*}[htb]
    \centering
    \includegraphics[width=\textwidth]{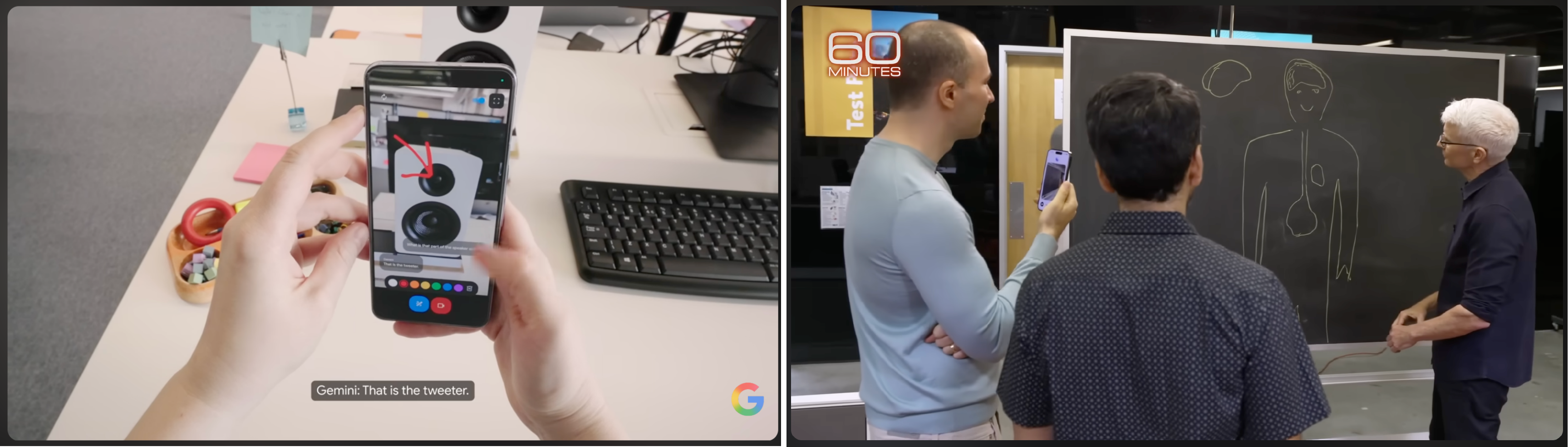}
    \caption{Screenshots from 2 of the 5 tech demo videos shared and discussed with educator participants during the workshop. On the left, a user demonstrates Google Project Astra \cite{projectastra} and how the MLLM is able to take input in the form of an annotated live video and describe what it sees (a part of a speaker). On the right, Anderson Cooper tries out OpenAI's GPT-4o, which uses live video input to critique Cooper's drawing of organs in the human body \cite{sixty_minutes}.}
    \label{fig:demo-videos}
\end{figure*}

Each 3-hour in-person workshops involved 3 participants and a research team of 4 whose members alternated between facilitating particular parts of the workshop and observing and note-taking. The workshops began with introductions and an overview of the agenda, followed by a 15 minute presentation introducing MLLMs and sharing product demo videos from Google’s Project Astra\cite{projectastra}, Google's Gemini\cite{gemini}, Apple Intelligence\cite{appleintelligence}, and ChatGPT-4o\cite{gpt4o}, some of which are displayed in Figure \ref{fig:demo-videos}. These videos showcase interactions such as conversing with a voice-based agent to play a map-based geography game (in which the model understands a user pointing at areas of a map in real time) \cite{gemini}, having the model interpret a user's drawing of the body to test their understanding of human anatomy \cite{gpt4o}, and generating alliterative poems based on an object the model sees \cite{projectastra}. The facilitator then led a group discussion in which educators shared their initial reactions, discussed how they would feel about students using such technologies for learning, identified modalities they found most exciting for educational contexts, and considered the subjects and learning goals where they believed these models might have the greatest impact. We then transitioned to a brainstorming activity where participants ideated on how MLLMs could be used to facilitate student learning in their own classrooms.

Next, a member of the research team facilitated a tutorial on Claude 3.5, the MLLM they would utilize for the remainder of the workshop. Participants followed along using the Claude iOS app on provided iPads and external keyboards (for details about why we chose to use iPads, please refer to Appendix \ref{sec:appendix-device-selection}). The presentation provided instruction on topics such as navigating the Claude interface, prompting best practices, applying ``custom instructions'' (system prompts), and sharing examples illustrating the Claude Artifacts feature (e.g., using an image of a vocabulary worksheet to create an interactive flashcard app for language learning). To give participants practice with writing and iterating on prompts, participants engaged in a small activity in which they took a picture of a page from a physical book and prompted Claude to describe it. The research team provided a variety of illustrated books, including storybooks and language learning books.

Participants then had 45 minutes to prototype a learning application of their choice using Claude. We provided a worksheet to scaffold their project planning process (shared in Appendix \ref{sec:appendix-project-planning-worksheet}). Throughout this activity, our research team provided support and answered questions the participants asked during their prototyping. Afterward, each teacher presented and demoed their projects to the group, outlining their design and iteration process. We ended with a group discussion where participants reflected on any changes to their perspectives on MLLMs after using it to build something of their own design. As compensation for their time, each participant received \$150.

\subsection{Data Collection and Analysis}
We collected audio recordings of the workshop activities (totaling 4 hours), screen recordings of participants' iPads (totaling 11.5 hours), video recordings of the educators’ prototype presentations (totaling 1.3 hours), and participants' Claude chat logs. Brainstormed ideas and project planning worksheets were captured via Miro boards used in these activities.

First, the four-member research team collectively categorized the brainstorming ideas teachers shared after watching MLLM demo videos but before they prototyped with a multimodal model themselves. We also examined the input and output modality types specified for teachers' ideas, enabling us to better understand which modalities teachers gravitated towards.

Next, recordings of prototype presentations and group discussions for all workshops (which took place after participants prototyped with Claude to build their own learning experiences) were transcribed and cleaned by the third and fourth authors. All members of the team read through the transcripts and collaboratively conducted reflexive thematic analysis \cite{braunUsing2006}, inductively identifying emerging themes around teachers' ideas for learning opportunities, challenges encountered when using MLLMs, and views on student use of MLLMs.

At this stage, our analysis revealed that educators differed in terms of whether they primarily saw student or teachers as end-users of MLLMs, with some prototyping learning experiences in which students use the model directly, while others imagined that teachers would generate content for students using the model. Based on this distinction, we selected one example of student-use and teacher-use to analyze more closely as a group. These two cases involve two different participants from our workshops that represent a dichotomy between two student age groups (elementary and high school), two different educational contexts (school-based learning and out-of-school learning in a museum), and student- versus teacher-use of the model. Additionally, the two teachers focused primarily on a single project goal for the 45 minutes, letting us more fully observe their debugging strategies compared to other participants who may have prototyped several ideas during the same time.

The second author performed open-coding of Claude chat logs from these two participants to identify types of multimodal inputs, prompts, and outputs from educators' interactions with Claude. We then triangulated using several additional data sources, including the educators' project planning worksheets, screen recordings of their interactions with Claude, video recordings of their presentation of their prototype, and audio recordings of our reflective discussion at the end of the workshop. We discussed these interpretations and codes as a larger 4-person research team in weekly research meetings over the course of two months to refine and reach consensus over our categorizations. We compiled visualizations of each participants' process that highlighted key moments in how educators engaged with the model to build their projects. These two perspectives are represented as vignettes in this paper.

After aligning on our analytical approach, the first author applied the same analysis to the remaining 10 participants. Additionally, this researcher examined all interactive artifacts created by Claude to verify the quality and accuracy of its response, which led to identifying instances in which the model generated inaccurate information. Collectively, our analysis enabled us to characterize the experiences and challenges encountered by all 12 participants.


\section{Results}

We organize our findings based on our two research questions, beginning with a description of the ideas educators had for learning opportunities using MLLMs (RQ1), followed by themes that emerged through educators' prototyping experiences and how they relate to challenges they anticipate for student use (RQ2).

\subsection{RQ1: What ideas do educators have for learning opportunities using MLLMs?}

In this subsection, we first report on the brainstormed ideas educators had using the full corpus of input and output modalities as shown in tech company demo videos we collectively watched and discussed. Since only a subset of these modalities were available in Claude 3.5, the MLLM the teachers prototyped with in our activity, we then separately report on the ideas teachers implemented using multimodal features available with this specific model.

\subsubsection{Brainstormed Ideas Using All Modalities}

In our initial brainstorms before participants prototyped with Claude, educators came up with 59 ideas for learning experiences using MLLMs. Of these project ideas, we categorized a subset of 47 in which teachers provided sufficient detail about how an MLLM would be used (for example, an idea such as `Cell Organelles - comparison to plant cells' did not provide a description of multimodal interaction and was thus excluded from our analysis). We identified seven categories: multimedia simulations and demos (45\%), personalization and accommodations (15\%), supporting student research (13\%), concept development and brainstorming (13\%), personalized tutoring (6\%), generating examples for student use (4\%), and enabling assessment (2\%)\footnote{Idea categories were not mutually exclusive.}. Table \ref{tab:ideas} presents these categories and a representative example of each, drawn verbatim from the educators' brainstorming Post-its notes. 

\begin{table*}[t]
\begin{center}
  \caption{Educator Ideas for Using MLLMs for Learning}
  \label{tab:ideas}
  \begin{tabular}{p{5.5cm}cp{8cm}}
    \toprule
    Category & \% of Total Ideas & Example \\
    \midrule
    Multimedia simulations and demos & 45\% & Audio interactive for students to find specific organs in the human body \\
    Personalization and accommodations & 15\% & Students could use it as a way to simplify information presented so they are able to access lessons in a modality that suits their learning needs \\
    Supporting student research & 13\% & Useful for plant identification, history, common uses\\
    Concept development and brainstorming & 13\% & Students can engage with text/audio for brainstorming reiterations of designs for design challenges \\
    Personalized tutoring & 6\% & Given a challenging math problem, students could use MLLMs to provide what information they would need to know to be able to solve the problem (i.e., vocabulary, step-by-steps, visual representations) \\
    Generating examples for student use & 4\% & It could be used to generate code and then students can figure out what is wrong with the code and how to fix the errors. \\
    Enabling assessment & 2\% & Students can use audio to describe their understanding of a text instead of writing it out on paper.\\
  \bottomrule
\end{tabular}
\end{center}
\end{table*}

Increasing student engagement through \textbf{multimedia-rich simulations and demos} was the most popular category, highlighting opportunities to move beyond text-explanations. Jessica emphasized the power of visuals: ``A picture’s worth a thousand words. I just feel like images and visuals -- they always add to whatever is text-based or just being said to you.’’ Both Laura and Michelle expressed how audio-visual media could be especially helpful for young learners that are still developing their ability to express ideas in text, while Rachel shared how students more broadly may have a preference for video or images given the media they regularly consume outside the classroom. 

The second most popular category was \textbf{facilitating personalization and accommodations}. 
Alex envisioned how he could use the model to rework generic material to use NYC-based details, better aligning with his students' lived experiences. Janet highlighted the importance of accommodating different learning styles for students who struggle with traditional assessments: ``All learners are not [able to] take tests on paper. I could see ways that students could be creative and use this platform [MLLMs] to provide what they might know using video or images.'' 


\textbf{Modalities Preferences}. We analyzed the input and output modalities represented in the educators' ideas at this stage of the workshop, finding that educators on the whole conceptualized multimedia (audio, images, and video) predominantly as output rather than input to the model. Specifically, the most popular output modalities were images (28\%) followed by audio (15\%), video (14\%), and text (12\%), while the most popular input modalities were text (27\%) followed by images (14\%), audio (10\%), and video (2\%)\footnote{Note that there were ideas in which teachers specified the output modality but not the input modality, or vice versa, which is why these percentages do not add to 100\%)}. One potential reason for the prevalence of text input and image output is that these modalities are more represented in popular use of LLMs and thus may be more familiar. Teacher George shared his excitement about alternative modalities: ``I was blown away by some of the stuff [in the demo videos]. I've only used text AI--I've done text to image or text to video, but never with the input being multimedia.''

\subsubsection{Multimodal Ideas after Prototyping with Claude}

Next, we summarize the ideas teachers prototyped using Claude 3.5, a model with support for image and text input and text and interactive code-based output.

\begin{figure*}[htb]
    \centering
    \includegraphics[width=0.9\textwidth]{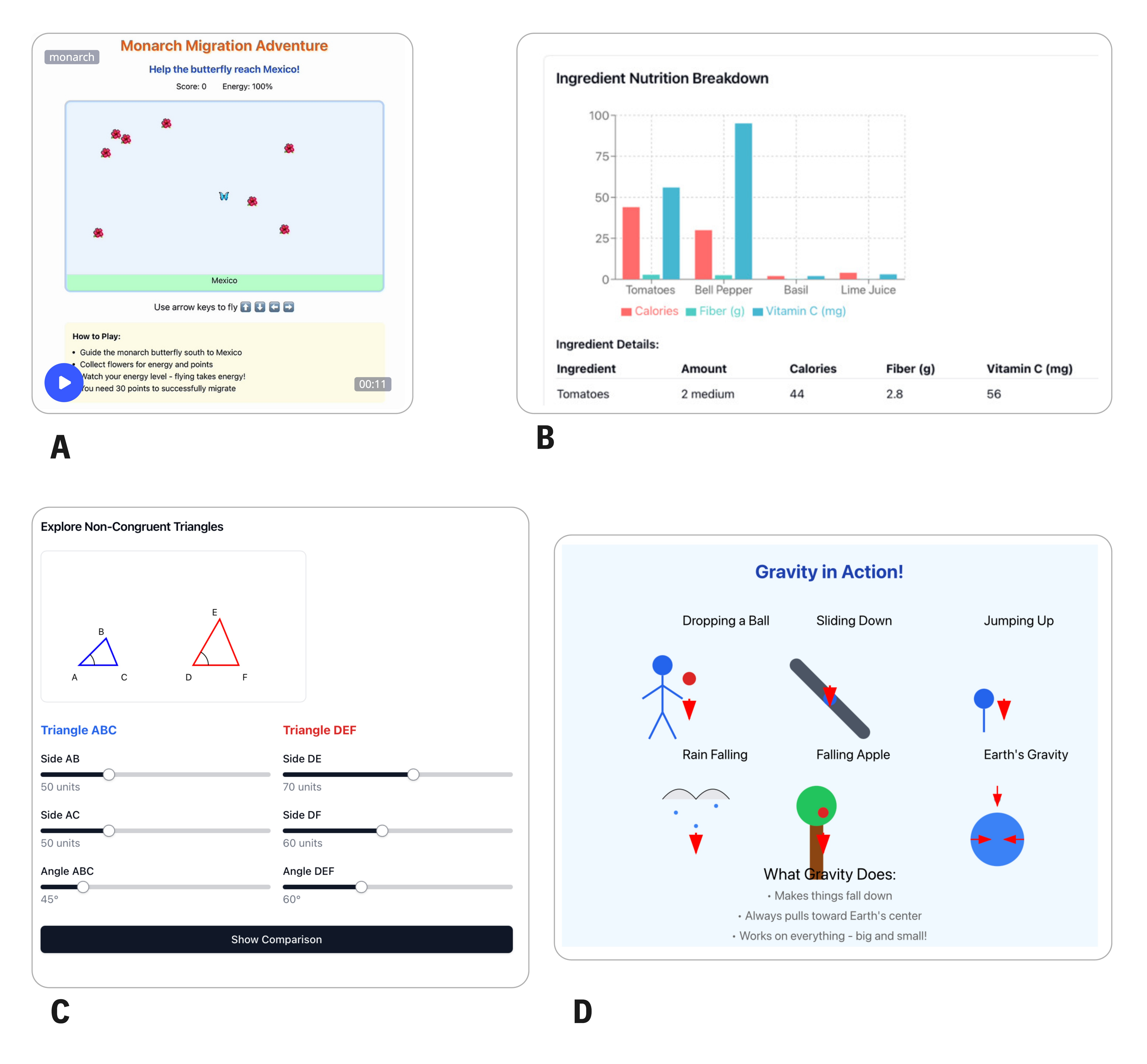}
    \caption{Interactive artifacts creates by workshop participants including A) a monarch-migration-themed game, B) a data visualization showing the nutritional breakdown of ingredients in a recipe, C) an explorable interactive for testing triangle congruence, D) a simple illustration of gravity acting in a playground.}
    \label{fig:interactive-projects}
\end{figure*}

\textbf{Types of Projects Created} Teachers created a total of 23 projects, with each participant creating anywhere from 1 to 4 projects. All but one educator incorporated image input, and all participants created some form of interactive code-based output. This was notable since during the initial brainstorms before using Claude, only one teacher brainstormed interactive outputs. The interactive projects created (summarized in Table \ref{tab:participants}) included games (34\%), data visualizations (17\%), explorable interactives (13\%), slideshows and flashcards (13\%), websites (9\%), and animations (9\%)\footnote{Interactive output categories were not mutually exclusive.}. Several project examples are displayed in Figure \ref{fig:interactive-projects}.

Educators' explorations of interactive artifacts exhibited how MLLMs have the potential to support teachers with limited programming experience to author custom learning applications. Even educators with prior experience may benefit from expediting the development process using an MLLM; for example, Mary shared, ``It's unreal how much I created in the 10 minutes [during the workshop] versus the hours it takes me to create them myself.’’ This result may be especially notable given that prior work identified how time constraints often prevent K-12 teachers from preparing and using multimodal content \cite{yi2015teachers}.

\textbf{Use of Image Input}. Given that image input is the primary multimodal feature available with the Claude 3.5 model, we provide additional detail about the ways in which teachers utilized this affordance. Teachers used the MLLM for \textbf{identification and comparison} of image content, prototyping applications in which students could ask the model to explain what was in an image or compare items in different photos. Two math teachers used image input to \textbf{provide hand-written mathematical content} such as geometric diagrams and algebraic equations. Photographs of diagrams (including a map showing differences in sea levels) were used alongside prompts requesting the MLLM to \textbf{transform static content into interactives} students could use to explore and reflect on the data. In other instances, images \textbf{provided local context} so that the model could tailor its responses to address a student's community (such as a particular neighborhood in New York City). One teacher used images of \textbf{anchor charts} (visual learning tools that link key concepts with images commonly used in classrooms) as reference images to guide the model's response, as shown in Figure \ref{fig:interactive-projects}B. Finally, one teacher provided images of characters from a children's book as a way for the MLLM to \textbf{reuse existing characters} in its generated output (a game shown in Figure \ref{fig:interactive-projects}C).  

In summary, we observed that images as input provided opportunities for personalization (e.g., the reuse of characters from a book, providing local context), supporting faster input of context compared to written text prompts (e.g., hand-drawn diagrams, anchor charts), or providing input in instances where users are not expected to have the vocabulary to construct a text-based prompt (e.g., identifying an item in an image, especially for children in the early stages of literacy).

\begin{figure*}[htb]
    \centering
    \includegraphics[width=\textwidth]{figures/results/image-input-types.pdf}
    \caption{Examples of image input from workshop participants' projects. A) Hand-written mathematical content, B) Anchor chart found on the web that relates facial expressions to names of emotions (created by Little Lucky Learners), C) Illustrations from a children's book (top) used to instruct the model to reuse existing characters in an interactive game (shown as simplified vector graphics in the matching game below). (Images from the children's book \textit{El pequeño pez blanco} by Guido Van Genechten)}
    \label{fig:image-input-types}
\end{figure*}

\textbf{Students vs Teachers as End-Users of MLLMs}. The projects educators built reflected two envisioned use cases: 1) learning experiences where \textit{students interact with the MLLM} directly, and 2) scenarios where \textit{teachers interact with the MLLM} to construct personalized content for student use. This difference was closely tied to the age of the students the educators' worked with. All educators who imagined students using the MLLM directly (25\% of the workshop participants) work with high-school students, who are more likely to have stronger abilities to prompt and constructively iterate on model output. In contrast, elementary and middle school teachers largely shared how younger students would need more direct guidance and support, as evaluating and refining model output requires a high level of critical thinking and persistence that they anticipated would be challenging for young audiences. To more fully illustrate the difference between students and teachers as end-users, we share two vignettes exemplifying these contrasting approaches.

\begin{figure*}[htb]
    \centering
    \includegraphics[width=\textwidth]{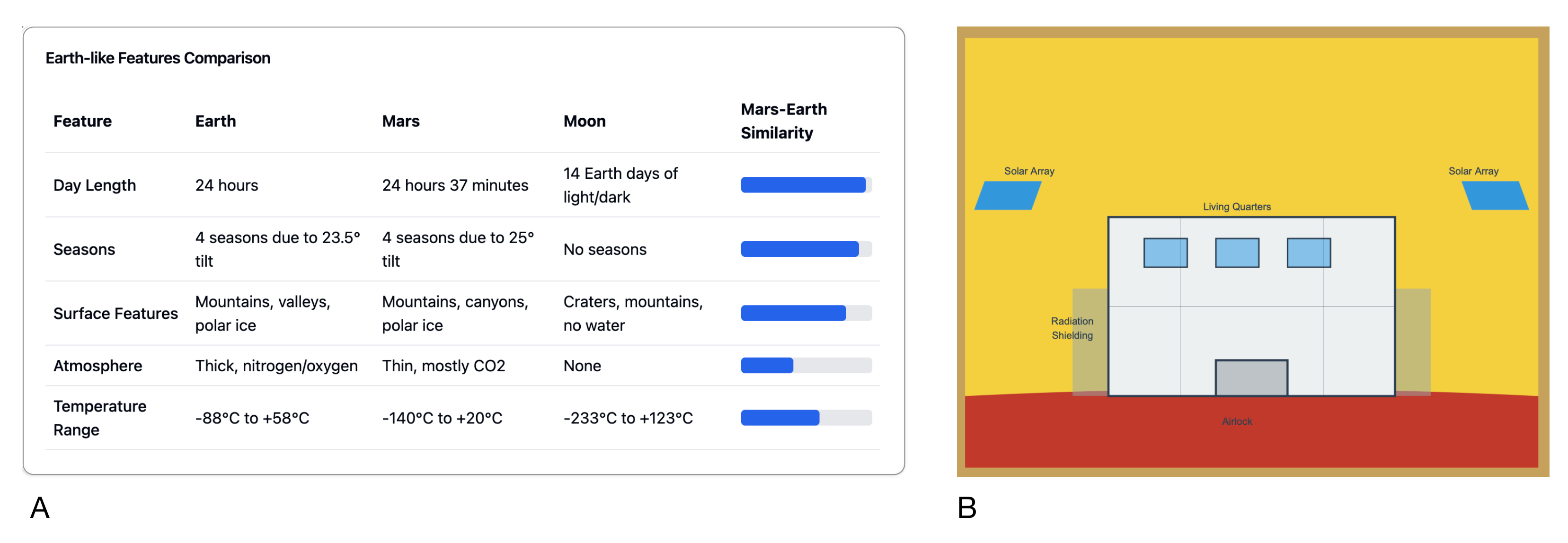}
    \caption{Artifacts created for Jessica's Space Research project, which include a chart outlining differences between life on Earth, Mars, and the Moon (A), along with a diagram of what a Mars habitat might look like (B).}
    \label{fig:case-study-2-artifacts}
\end{figure*}

\textit{Student as MLLM End Users: Supporting Student Research}. Museum educator Jessica drew from a research-based design challenge she already uses in a 2-week STEM summer program for high school students. For this challenge, students design a habitat for humans to live on Mars. In her experimentation with Claude, she put herself in the position of a student, imagining how they might use the model to do research necessary to support their design. She began by prompting the model to \textit{``compare and contrast the environments of the Moons and Mars.''} In response, the model generated a chart outlining key differences between Earth, the Moon, and Mars as shown in Figure \ref{fig:case-study-2-artifacts}A. She remarked that she was happy that the table served to synthesize facts to actively support comparison, a process she expects her students to engage in to make informed design decisions. She then prompted the model to \textit{``Show me a simple design for a mars habitat’’}, receiving a simple generated SVG image (shown in Figure \ref{fig:case-study-2-artifacts}B) that depicted solar arrays and radiation shielding. While the illustration was rudimentary, she saw it as an opportunity for students to modify and improve on the output, rather than using it as is. She imagined students reacting to the generated image by saying ``OK, this is a good start, but I can make it look better.’’ That is, she saw the limitation of Claude's simplified SVG graphics as providing a starting point that would necessitate students' own creativity to improve and make their own.

\begin{figure}
    \centering
    \includegraphics[width=\columnwidth]{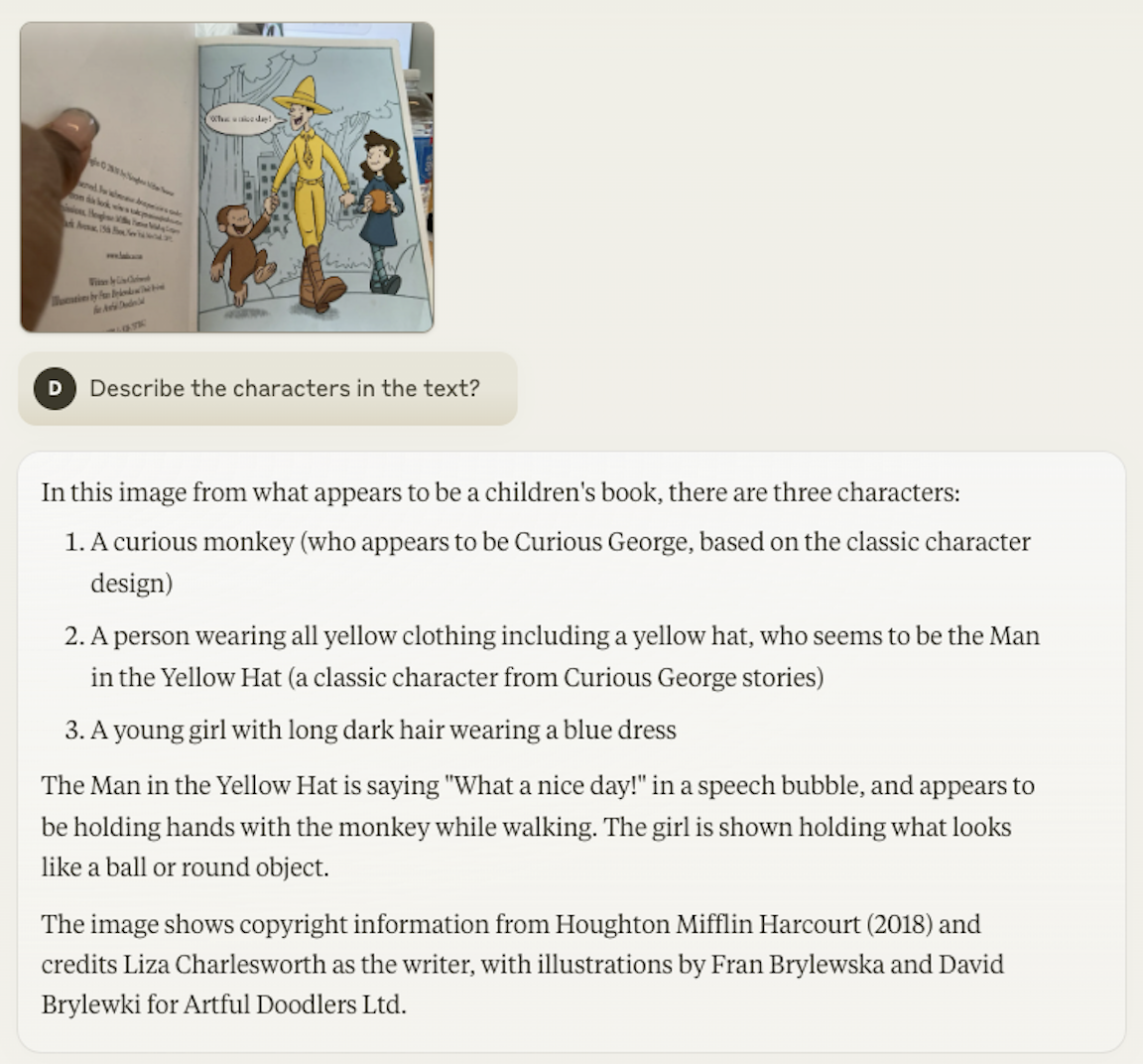}
    \caption{Claude interpreting the first page of the book from \textit{Curious George Follows that Hat}.}
    \label{fig:curious-george}
\end{figure}

\begin{figure*}[htb]
    \centering
    \includegraphics[width=\textwidth]{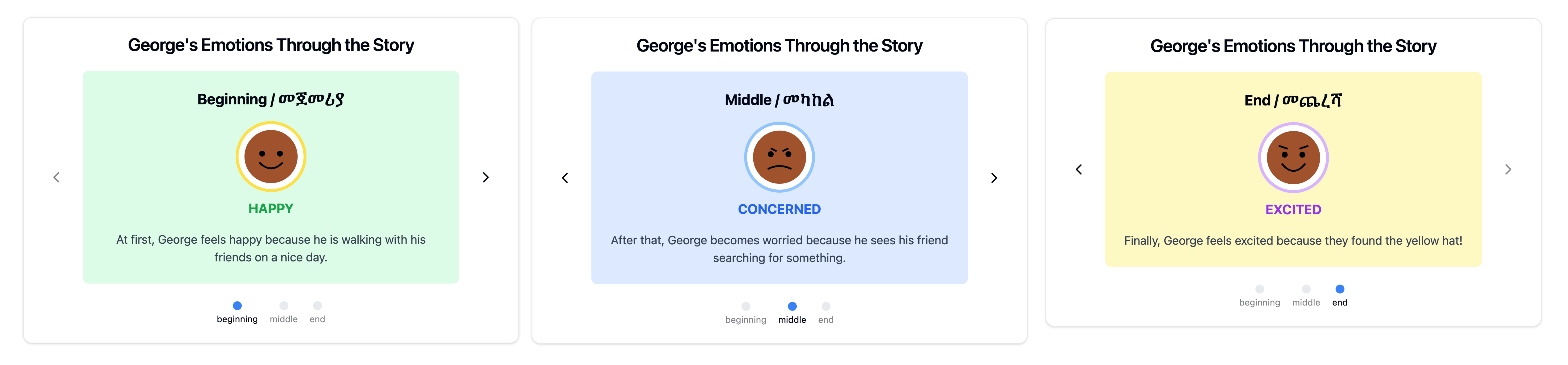}
    \caption{Interactive slideshow created by Jessica and Claude for exploring the beginning, middle, and end of the book.}
    \label{fig:case-study-1-artifact}
\end{figure*}

\textit{Teacher as MLLM End Users: Creating Interactives for Student Use}. Inspired by reading activities with her fourth grade students, Janet created a project using the illustrated book ``Curious George Follows that Hat.'' She first explored the MLLM's ability to interpret illustrations by uploading an image of the first page of the book. In response, the model identified the main character as Curious George and described the appearance of other characters in the scene as shown in Figure \ref{fig:curious-george}. She then proceeded to create a reading guide for supporting students learning to use transition words to describe a character's emotional journey throughout the story. In her Claude chat, she added images of pages from the middle and end of the book, as well as web-sourced anchor chart images showing transitional words and emotions, prompting the model to \textit{``create an interactive application with visuals representing the emotions of beginning, middle, and end''}. The resulting generated slideshow provided a summary of Curious George's emotions throughout the book as displayed in Figure \ref{fig:case-study-1-artifact}.

Janet also attempted to incorporate multimodal audio input and output to support a non-native English speaker in her class. However, when she found that Amharic was not supported in Claude's text-to-speech input, she instead used the model to incorporate Amharic translation to the slideshow's text. Over the course of 55 minutes, Janet repeatedly refined the model's response to better match a fourth-grade reading level over 8 different system and user prompts, ultimately steering the model to transition from lengthy text responses into interactive slides each containing a single short sentence. Reflecting on her experience, Janet explained that while she originally imagined her students using the conversational interface directly, the amount of effort required for her to refine the model output led her to view the MLLM primarily as a teacher-facing tool with which she could more fully ensure alignment with her teaching goals.

 \subsection{RQ2: \rqtwo}

Here, we describe several themes that emerged from analyzing challenges participants' faced when prototyping with an MLLM, combined with our analysis of the quality and accuracy of generated artifacts. These experiences directly shaped how our participants perceived the relevancy and potential feasibility of using MLLMs with youth.

\subsubsection{Critically Assessing Generated Content}

\begin{figure*}[htb]
    \centering
    \includegraphics[width=\textwidth]{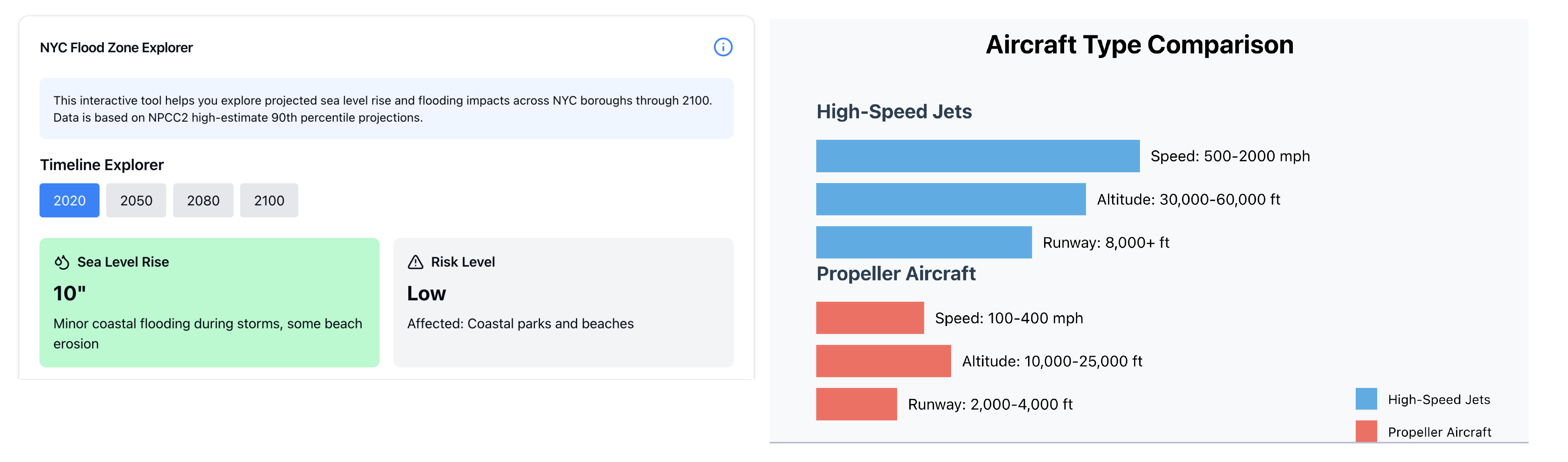}
    \caption{Artifacts from Claude that incorporate generated data. While Claude cited its source in one of its Artifacts (top left), other generated data visualizations did not cite their sources, raising the issue of how students and teachers can verify data presented in MLLM generated media.}
    \label{fig:data-generated}
\end{figure*}

Teachers frequently used the MLLM to request and visualize information, which produced artifacts including tables and graphs of data as shown in Figure \ref{fig:data-generated}. But in all but one instance, the LLM did not cite its sources. Teachers Rachel and Alex expressed the importance of including references to help students ``fact-check'' and ``look deeper'' in the case they want to learn more. However, other teachers were surprisingly uncritical of Claude's output. For example, teacher Michelle shared that she was impressed that, ``[Claude] allows [students] to see actual reliable data they can use to answer their inquiry question,'' and it was only when a researcher questioned how she knew the data was reliable that she realized that the sources were not provided. Similarly, Jessica, who was excited about having students use MLLMs for research, did not appear to question the validity of the data while using the model herself, but later in the group discussion agreed that references would be helpful for \textit{students}. 

The MLLM commonly generated collections of facts about various topics, including facts about animals, nutritional information, and planets. Our analysis of facts generated in educators' artifacts found that the facts themselves were accurate.  However, in several instances where the MLLM tried to generate diagrams related to physics and mathematics, it produced inaccurate representations, such as mislabeling the direction of gravity and normal forces and not indicating an open circle for a greater than value on a number line. 

Teacher Patricia, whose experiments with the MLLM generated an inaccurate chain reaction animation (in which dominoes fell one after another without touching one another), saw incorrectly generated content as a potential learning opportunity. She described how she would ask her students to identify the incorrect behavior and try modifying the prompt to fix the visualization. While this presents an optimistic view for engaging students in fixing model errors, these errors more broadly pose risks for misinformation if left unidentified. In educational contexts where trustworthiness of data and practices around citing data sources are crucial, there is a need for models to cite their sources, particularly \textit{within} generated artifacts (not solely in the accompanying text description), especially if teachers envision the models being used to support student research. While hallucinated data is a concern with any LLM, inaccurately generated data with \textit{MLLMs} specifically may pose different risks, as they may be presented in polished visualizations and figures that give the impression of precision and correctness. Additionally, given some educators' uncritical use of the data, providing teachers and students with instruction on how to assess model output quality and accuracy may be particularly important in classroom use of MLLMs.

\subsubsection{Generated Content as Intermediaries}

In our initial discussions, fear of student overreliance on LLMs as a shortcut for answers was the most common concern, with educators fearing that LLMs would ultimately hinder students' original thinking. However, after gaining direct experience prototyping with MLLMs themselves, teachers described how their notions of the role of AI in education expanded, especially in using LLMs to support larger learning goals. Educator Rachel shared how ``It's [MLLM] not just giving you an answer – it kind of gives you a lot of options'' which echos Karen's sentiment that MLLMs can support idea building: ``It's not just students looking up answers...it's giving. It can generate ideas [for the students] to then do something on their own.'' These educators described how generated LLM content can create what we describe as \textit{intermediaries}, or artifacts that may help students consider alternatives, rather than be used to directly generate solutions.

In the context of multimodal generated content, Patricia summarized this by stating, ``We could definitely use this [generated] visual to help with [student] understanding – not just putting in a question to get an answer, but putting in a question to get this image to give you that broader view of what you were learning about.''  Alex reiterate this idea by sharing how interactive output can ``show interactive ways of learning, whether it's graphs [or] simulations...it's a different way to see and answer questions in a creative way.'' These statements exhibit sentiments about how models might support students with conceptualizing a design space through providing visuals that enhance their understanding and serve as just one piece of how they communicate their learning.

The notion of generated content as intermediaries appeared to stem from the incomplete nature of what Claude was capable of creating and the effort required to iterate with the MLLM to get closer to one's goal. For example, as shown in our vignette of educator Jennifer's process, the rudimentary vector graphics generated by Claude were seen as opportunities for students to use them as first drafts that they modify and extend, rather than use as is. Teachers Mary and George also described how the MLLM rarely produces the design you expect from an initial prompt, and that it is through the continuous refinement with follow-up prompts that a learner would crystallize design requirements and goals. They discussed how if students were given the opportunity to use MLLMs for design projects, they would expect them to be able to `show their work' by describing the series of prompts they used to create an artifact, aligning with iterative research and engineering design process.

\subsubsection{Reflections on Feasibility of Use} 
Teacher concerns about MLLMs reflect prior findings about use of AI technologies more broadly \cite{braaten, nazaretsky2022teachers}, such as barriers to access to AI technologies in classroom settings (for example, our workshops utilized Claude Pro, which has a monthly \$20 subscription fee that would be not be scalable for entire schools) and a lack of clear policy around model regulation and access \cite{dangol2026relief, xiao2026teachers}. Since all teachers in our study work in New York City, the fact that ChatGPT was at one point banned by the NYC public school system \cite{rosenblatt2023nyc_chatgpt_ban} was brought up as a sign of the uncertainty of educational guidelines around the use of AI models. 

When it came to model responses, all educators found that the default generated text explanations were lengthy and needed to be highly tailored (through system prompting or follow up prompts) to be more digestible to students with varying literacy levels. Given the amount of effort educators put into steering model output via iteratively refining system prompts, it remains an open question how these models might be more easily adapted to flexibly serve different age groups in terms of their knowledge of the material and their vocabulary level.

Finally, we found a significant gap between tech company demos and what publicly available models are actually capable of, which led to inflated user expectations. We see the need for level-setting not only what modalities are available, but at what quality these MLLMs can produce. One such example is how code-based images generated with Claude are much more simple and abstract than diffusion-based generated images. While Jessica (Vignette 2) appreciated Claude's simple graphics, other educators encountered overly abstract images they deemed to be detrimental to the learning experience. For example, a generated game about organs in the human body depicted organs as abstract shapes that did not resemble their appearance in real life, which the teacher felt would lead to student misconceptions. Beyond images, teachers wanted audio input and output to support English language learners, but audio was not supported using Claude at the time of our study beyond speech-to-text input. Ultimately, the gap between polished tech videos and actual capabilities became tangible when experimenting with these models hands-on, pointing to the need for better user education materials and opportunities that engage end-users with practically applying these models to understand their capabilities and limitations.
\section{Limitations}
\label{sec:limitations} 

The perspectives shared in our work necessarily reflect the backgrounds of our participants, which come from an urban setting based in the United States. To the best of our abilities, we tried to capture a range of perspectives by recruiting from teachers serving a wide range of students (elementary, middle school, and high school students), subject areas, and in- and out-of-school contexts. Additionally, the participating educators might be more optimistic about AI technologies than the average teacher as they elected to participate in our AI workshop in their free time. Future work examining perspectives from a broader range of educators may, in particular, bring forth more critical perspectives, particularly from AI skeptics.

Despite showing demo videos that incorporated audio input and output at the start of the workshop, we saw few examples of these inputs in participants' brainstormed ideas. We were interested in whether learning applications leveraging audio input might not have been strongly represented because of privacy concerns around child voice-based input in classroom or group settings, but we were unable to discuss this as we only recognized after the workshops that few audio-input ideas were brainstormed. Additionally, Claude 3.5, the MLLM teachers prototyped with, did not support audio and video input and output at the time of our study (November 2024), so we did not have the opportunity for teachers to experience these capabilities directly. With Claude now supporting Voice Mode as of May 2025, future work might expand on educator impressions of audio-based modalities.

Our work focuses on teachers' initial impressions and experiences prototyping with MLLMs. We see future work expanding these initial engagements to study extended use of MLLMs in classrooms and learning contexts with students. This would be a necessary next step towards unpacking the contextual factors that shape real-world integration.
\section{Discussion}
\label{sec:discussion}

Our study provides an early lens into teacher perspectives on educational opportunities applying MLLMs. As multimodal features are just beginning to be integrated into commercial tools, learning from end-user perspectives can help both designers and researchers better anticipate how these tools might be extended and adapted to best support educational uses cases. Here, we reflect on the implications our findings have on applying MLLMs to learning experiences that align with educator values.

\subsection{Code-Generated Multimedia Versus Multimodality}
In our workshops, teachers designed and prototyped interactive applications built with generated code using Claude Artifacts. One surprising result of Claude Artifacts is that although the output is technically text-based (code) and not multimodal, the interactive applications teachers prototyped using Artifacts captured the spirit of what might be created using a multimodal model, such as images (in the form of simple vector graphics), animations, and visualizations. These types of artifacts align with the Cognitive Theory of Multimodal Learning \cite{mayer1997multimedia, mayer2002multimedia, mayer2024past} and the value that visuals bring beyond text alone, suggesting that some \textit{visual} content proposed by Bewersorff et al. \cite{bewersdorff2025taking} may be possible to create today using previews of generated code rather than truly multimodal output. This in turn may make it more accessible for visions around multimedia learning to be realized today, though there are still open questions about aligning the resolution of generated content with end-user's expectations, as we saw with the limitation of code-based graphics that were considered too rudimentary for some of the educators' needs.

Yet, a notable distinction between the ideas educators brainstormed and the ones presented in tech demo videos discussed during the workshops is a lack of real-time interaction made possible with video and audio input and output. For example, product videos often presented scenarios in which users converse with the MLLM using live video input and are able to verbally pose a question to the AI agent and receive an auditory generated explanation. Audio and video capabilities are not widely available in popular chat interfaces like ChatGPT and Claude and thus were not possible for teachers to experiment with at the time of our study. As these features become available, future work can consider in what ways these modalities might shift what educators and students imagine as useful educational applications of MLLMs.  However, there still remains a question of how to best steer model output to adapt to learners of varying knowledge and literacy levels, as we describe next.

\subsection{Unpacking the Needs of Teacher- Versus Student-oriented Use Cases of MLLMs}

An open question for further research is how these AI tools might be better adapted to serve distinct teacher- or student-oriented needs, as illustrated by our two vignettes. The majority of our participants focused on using the MLLM for teacher-created content for student use, as opposed to designing learning experiences in which students interact with the MLLM directly, distinguishing our findings from prior work on how older undergraduate students envision \textit{themselves} using MLLMs for learning \cite{prasad2025exploring}. Part of the divide of teacher versus student-oriented use cases came from how elementary and middle school teachers (compared to high school teachers) anticipated challenges young people might have with effectively interacting with an MLLM, including questions about an MLLM's ability to handle varying student literacy levels, whether young students have the critical thinking skills needed to evaluate model output, and if young learners have the persistence required to iteratively interact with an LLM in the case its output is not aligned with their requests. In other words, even if the video and audio capabilities of MLLMs were accessible to end-users today, it is still uncertain how these models can flexibly adapt to younger audiences who can vary significantly in their ability to effectively communicate their goals and level of understanding to an AI agent. Thus, an open question is how we can either adapt models to provide more appropriate explanations for younger audiences, or empower educators and students as end-users with suitable levers for steering model output themselves.

\subsection{Fostering Students Generating Supporting Material Rather Than Solutions}

For student-oriented use cases, educators were most excited about opportunities for MLLMs to engage students in developing their own perspectives and content, rather than more traditional question-and-answer interactions shown in the tech demos. For example, educators imagined scenarios in which MLLMs can generate intermediaries such as data visualizations that could help spur students' critical thinking towards more general project goals. Rather than using LLMs for generating a solution, AI was imagined as a way to ``help to get ideas and expand on ideas,'' offering ``more ways to approach one simple thing'' than a single answer, as described by educator Alex in our workshop.

While in initial discussions, many teachers expressed concerns about students using LLMs as a shortcut for answers, after prototyping, teachers reflected upon how MLLMs can support student exploration and research. Multimedia output such as visualizations could synthesize and present what would normally be text-based responses, enabling students to see patterns and make comparisons. Further, we saw in the example of Jessica's project-based activity (Vignette 2) that even when the MLLM generated low-resolution SVG images, this generated content was seen as a skeleton for students to build on top of, allowing them to integrate their own perspectives to fully customize their work. The perspective of using MLLMs to support students exploration of ideas, comparison of alternatives, and building off of generated content represents a way of learning with LLMs that go beyond more rote learning interactions (e.g., testing one's knowledge or question-and-answer explanations) that are commonly presented in tech demo videos. In other words, the MLLM may potentially serve as an ideation and research tool, but students need to be prepared to critically consider the validity of generated output and ultimately make their own choices on how to incorporate these pieces of knowledge into their own work, emphasizing student agency and decision making.

\subsection{Refining Artifacts as a Pathway Towards Iterative Design}

Many of the tech demo videos we reviewed and discussed in our workshops showed a learner-agent interaction in which the AI largely serves to provide explanations to questions. In contrast to this approach, the envisioned use cases teachers imagined and prototyped were largely focused on using the MLLM as a \textit{design tool} to create multimedia artifacts. When teachers described how they envisioned students engaging in this process, they raised how the MLLM rarely understood their own requests in its first attempt, and that iterative prompting was required to articulate one's design goals and refine model response to get closer to those goals. If students were to use these models themselves, the learning would stem from the student building their capacity to crystallize their goals when interacting and iterating with the model (in contrast to a question-answer interaction where iterating may not be inherent to the process). 

Design iteration was even brought up in instances in which students may not be using the model themselves, such as group debugging of artifacts as described by teacher Patricia. In cases like this, the educator could serve as a mediator, posing questions to help students consider whether the representation shown in generated content is correct, and working together with students to try reprompting the model to refine its output in response to errors.

While reprompting is a process that is a common interaction pattern with any LLM, we see MLLMs specifically as providing a unique pathway given its visual and potentially interactive output –– that it may serve as a quick means to \textit{see} the model's assumptions about one's user goals and provide an opportunity to test and reframe those assumptions through follow-up prompts. Prior work from \cite{keppler2025making} identified \textit{make for me} as prompt mode in which educators employed LLMs to generate content (such as quizzes and problems for student assignments). We see the creation of artifacts with MLLM as a type of \textit{make for me} use case, but because the model makes a range of assumptions about how an multimedia artifact should look and feel, make for me is rarely a one-shot process and instead requires design iteration. We see this as an exciting direction for future work to explore how this iterative design practice might be enacted by students and teachers in extended classroom contexts.
\section{Conclusion}
\label{sec:conclusion}
Multimodal LLM product videos from major tech companies like Google and OpenAI commonly paint a vision for how this technology will shape the future of learning, demonstrating how models can flexibly gather user context from video, audio, and images to help users understand their world. However, it is not known whether this vision of learning aligns with how educators imagine meaningful student learning experiences using multimodal models. In this work, we provide a first look into teacher perspectives and experiences using MLLMs to design their own learning experiences. In workshops with 12 K-12 educators from both in- and out-of-school learning environments, we engaged educators in discussions about the visions of learning presented in product videos of unreleased multimodal features, while also providing a hands-on experience for teachers to test a subset of multimodal features using Claude 3.5 to prototype their own learning experiences. We found that teachers of younger students (elementary and middle school students) were more likely to see MLLMs as tools that \textit{educators} can use to design content for students, while teachers of high-school-aged students saw opportunities for \textit{students} to engage with MLLMs to support their own research.  Further, through creating generated code-based interactives, or Artifacts, using Claude, teachers were able to realize part of the vision of multimedia-rich learning content, even without full multimodal capabilities, which presents an exciting opportunity for teachers today to begin engaging with multimedia creation with LLMs beyond text. However, outstanding challenges around critically evaluating generated content and understanding the capabilities and limitations of these models highlight the need for further work around professional development to support future integration of these technologies into classrooms and other learning contexts. We hope that these results may inform how designers of end-user AI tools take into account diverse learning goals and opportunities leveraging MLLMs.

\printcredits

\bibliographystyle{cas-model2-names}

\bibliography{cas-refs}

\appendix
\section{Appendix}

\subsection{Device Selection}
\label{sec:appendix-device-selection}
We considered whether we wanted workshop participants to interact with Claude on mobile, using a tablet, or on a laptop computer and decided on tablet-based interaction with external keyboards provided for the following reasons: 1) since text was the dominant form of providing input (even if supplementary images or media could be uploaded), we wanted users to be able to easily type requests with a physical keyboard; 2) we wanted to support users capturing pictures using a mobile device, which is more flexible than using a laptop webcam; 3) tablets would enable more screen real estate for evaluating model responses than a mobile phone, especially for interactive Artifacts; 4) tablets are more commonly used in educational contexts that mobile phones.

\subsection{Project Planning Worksheet}
\label{sec:appendix-project-planning-worksheet}
We provided a worksheet to scaffold participants' prototyping process, which included sections for specifying their learning goal, what their system prompt might be, what they could provide as input to the model, what their text prompt could be, and what they expect the model output to be.

\begin{figure}[htb]
    \centering
    \includegraphics[width=0.5\textwidth]{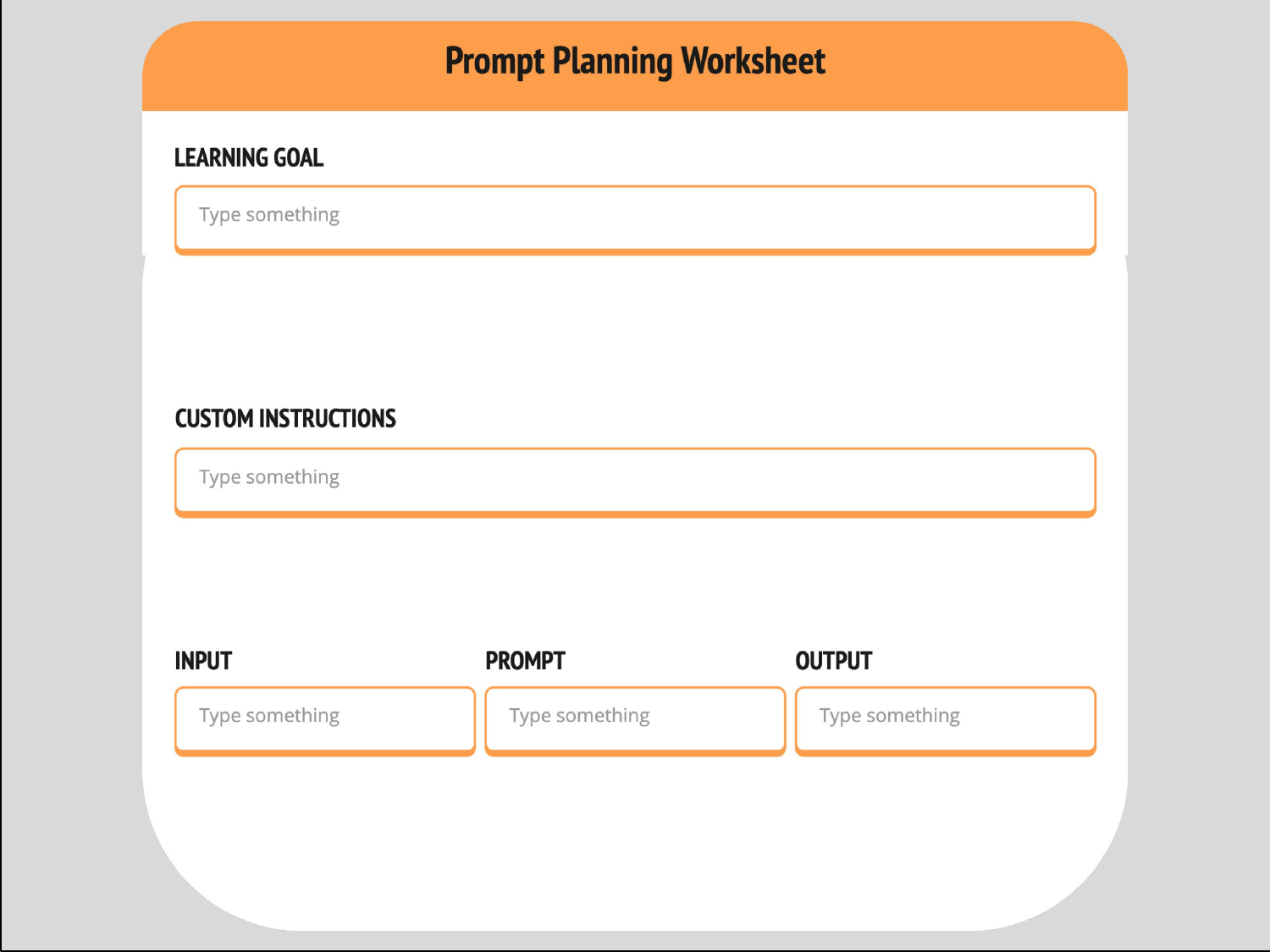}
    \caption{Prompt planning worksheet filled out by participants prior to freeform prototyping.}
    \label{fig:methodology-prompt-planning}
\end{figure}





\end{document}